\documentclass[12pt]{article}
\usepackage{epsfig,a4}
\begin{document}
\def\slash{\not\!}
\def\sslash{\not\!\!}
\def\vecalpha{{\mbox{\boldmath{$\alpha$}}}}
\def\vecnabla{{\mbox{\boldmath{$\nabla$}}}}
\def\vectau{{\vec\tau}}
\def\vecp{{\bf P}}
\def\vecr{{\bf r}}
\def\barpsi{\bar\psi}
\def\d{{\rm d}}
\title{Nuclear phenomena derived from quark-gluon strings}
\author{
Henrik Bohr\\
{\it \small Department of Physics, B.307, Danish Technical
University,}\\{ \it \small 2800 Lyngby, Denmark}\\
 Constan\c{c}a Provid\^encia and Jo\~ao da Provid\^encia\\
{\it \small Departamento de F\'{\i} sica, Universidade de Coimbra,}\\
{\it \small  3000 Coimbra, Portugal}  }

\maketitle

\abstract{We propose a QCD based many-body model for the nucleus where
  the strong coupling regime is controlled by a three body string
  force and the weak coupling regime is dominated by a pairing
  force. This model operates effectively with a quark-gluon Lagrangian
  containing a pairing force from instantons and a baryonic string
  term which contains a confining potential. The unified model for
  weak and strong coupling regimes, is, however, only consistent at
  the border of perturbative QCD. The baryonic string force is
  necessary, as a {stability and} compressibility analysis shows, for
  the occurrence of the phases of nuclear matter. The model exhibits a
  quark deconfinement transition and chiral restoration which are
  suggested by QCD and give qualitatively correct numerics. The
  effective model is shown to be isomorphic to the Nambu-Jona-Lasinio
  model and exhibits the correct chirality provided that the chiral
  fields are identified with  the 2-particle strings, which are
  natural in a QCD framework.}


\section{Introduction}

It has long been an important task for nuclear physicists to
accommodate the most accepted fundamental quark interaction, QCD,
in a field theoretical setting. This is because the nucleus
constitutes the most important observational bound state system
governed by the strong interaction. Furthermore, since the
nucleons, the elements of the nuclear bound state, are believed to
consist of quarks for which the rigorous strong interaction field
theory applies, it is natural to try to describe the nucleus in
terms of quarks (and gluons) beside nucleons. The present paper
therefore aims at explaining nuclear properties and structure in
terms of quarks and gluons. Still, one could be skeptical about
using quarks with no direct observables in the description of
nuclear physics that is very much based on experimental data.
However, since the occurrence of strangeness excitations in
hypernuclear physics needs to be explained by isospin or flavour
quantum numbers that naturally come out of a quark model, there
are good arguments for including quarks in nuclear physics.

There has been the obstacle of having a strong quark coupling
constant above unity in the interesting range of the strong
interaction which applies to the bound state and resonance
phenomena. This strong coupling regime is {characterized} by low
momentum transfer and corresponds to the region where quarks are
believed to be confined within the hadrons, contrary to the high
momentum transfer behavior where the coupling is small and quarks
are believed to be free. A coupling constant above unity in the
region of the strong interaction which is interesting for nuclear
physics will basically render a perturbative description of the
strong interaction field theory useless. We shall therefore use
the trick of putting in quark clustering by hand, so that
nucleons are treated as elementary particles, but quark structure
is taken into account in the interactions by assuming that the
mass of the nucleon  depends on the medium through the local
values of the fields $\sigma,\,\vec\pi$. Once that is done we
shall try to perturb the energy around the solution so obtained
and else try to appeal to similar phenomena in other areas of
physics (i.e. fluid dynamics) where solutions can be migrated to
the nuclear system.

In the beginning of the paper we shall introduce the quark
description
of the nuclear system and then propose a many-body Lagrangian
for the nucleus. Then we shall go on with a description inspired
on
fluid dynamics and try to derive phases that could be relevant
for
nuclear physics.

\section{A quark model for nuclear structure}

In the following chapters we shall propose a general Lagrangian for the nuclear
system in terms of quarks and gluons, and based on an interaction
like that of gauge theories
in the strong interaction version of quantum chrome dynamics, QCD.

Basically the full Lagrangian is like a chiral model but we shall
in the following list the various ingredients of the energy
functional.

\subsection{The pairing force}

A starting point for a QCD model for the nucleus in terms of quarks and gluons could be a kind of
a bag model in analogy to the well-known MIT model \cite{2)} for baryons and mesons but bearing in
mind that the nucleons themselves are MIT bags. In the usual setting of the MIT model there
is a strong  influence of phenomenological aspects that we eventually will try to derive with field theoretical interaction terms.

The generic form of the MIT bag model operates with a back
pressure that ensures the quarks to be confined to the bag and a
Coulomb force between the quarks. In the case, as here, where the
entire nucleus is to be considered as an MIT bag with quarks
moving within the boundaries of the bag, the bag pressure is again
an infinite potential well that keeps the quarks confined to the
nucleus. The quarks are, as usual, fermions with flavor and color
charges. As for the
interaction between the quarks, a crucial pairing force between the isospins 
ensure the total
isospin to be zero. In the next subsection we introduce a baryonic string force that
is responsible for
nucleon formation and actually also for the bag pressure.

The pairing force which has been
derived from instanton contributions in QCD field theory
\cite{hooft} is able to account for important features of nuclear structure.
For the one-instanton contribution the
pairing force is a truly 2-body force. Its expression is:
$$P_2 = g\bar\psi_i (x) (1+i\gamma_5) \psi_{i'} (x) \bar\psi_j (x) (1+i\gamma_5)
\psi_{j'} (x) \varepsilon_{i'j'}\varepsilon_{ij}$$
where $\varepsilon_{ij}$  anti-symmetrizes the isospin indices
$i,j$.

In the color zero sector, this effective interaction can be
reduced to the Nambu-Jona-Lasinio form \cite{Klevansky92,PRS},
$\beta_1\beta_2(I_1I_2-(\vec\tau_1\gamma^5_1)\cdot(\vec\tau_2\gamma^5_2)).$
In that case the missing terms, which constitute the difference
between both expressions
$$
\beta_1\beta_2\left[(I_1+i\gamma^5_1)(I_2+i\gamma^5_2)(I_1I_2+\vec\tau_1\cdot\vec\tau_2)-
(I_1I_2-(\vec\tau_1\gamma^5_1)\cdot(\vec\tau_2\gamma^5_2))\right]\,.
$$
average out to zero in a mean field approach.

\subsection{The baryonic string force in the nuclear bag}

\begin{center}
\begin{figure}[t]
\begin{center}
\epsfig{file=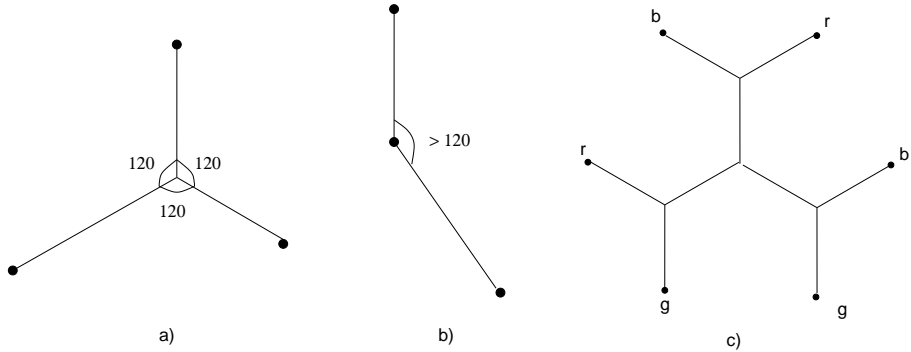,width=12cm} 
\caption{Baryonic string-configurations. (a) The angles of {the triangle defined by} a 3-quark cluster
are all less than $2\pi/3$. (b)One of the angles of {the triangle defined by} a 3-quark cluster
is greater than $2\pi/3.$ (c) A 6-quark cluster.  
This configuration 
is here included as a possibility but not  used in the full picture.
It will be discussed in a forthcoming paper.
\label{fig:fig1}}
\end{center}
\end{figure}
 \end{center}
In this 
subsection we discuss the many-body string force term.
We shall therefore introduce a three body string force to the Hamiltonian.
This force replaces the MIT bag pressure and is supposed to be
responsible for nucleon  bag formation. The string force is
proportional to the distance between the quarks being a very
dominating force, so perturbation theory with the string force as a
perturbation is not possible. On the other hand the strings will
triplet-wise neutralize the color leaving us with only dipole or
quadrupole moments. The string force is a qualified
guess to a configuration -- an ansatz.

We shall now specify the baryonic string configuration. We have
only limited information about nonperturbative string
configurations. Lattice QCD gives, in the strong coupling limit,
indication of color tube strings between two heavy quarks. To make
a consistent picture of the pairing and baryonic forces we are
staying within the border of perturbation theory, the strong
coupling constant, $\alpha_{QCD}$, being around 1.

We can, however, resort to the dual string theories for help concerning string parameters.
Here, the string 
interaction is proportional to the minimal length
$L$ of the string:
\begin{equation}
V_{st} = {1\over 2\pi \alpha'} L = \sqrt{\sigma} L
\label{2.1}
\end{equation}
where $\alpha'$ is the Regge intercept and $\sigma$ is the string
tension. In the sequel, to avoid confusion with the sigma field,
we write $\kappa=\sqrt{\sigma}$. The total 3-body force is then,
apart from color 0 projection operators,
\def\d{{\rm d}}
\begin{equation}
V_3 = \int \d x_1 \d x_2 \d x_3 \bar\psi (x_1) \bar \psi (x_2)
\bar\psi (x_3) V_{st} \psi (x_3) \psi (x_2) \psi (x_1)\,.
\label{2.2}
\end{equation}
The two possible \cite{4)} configurations between 3 quarks in
$SU(3)$ - QCD are shown 
in Fig. \ref{fig:fig1} when we require $SU(3)$ gauge invariance.
This means that quarks cluster in multiples of 3. Within our approximation
we only consider two and three body forces arising in the color neutral configurations,
Fig. \ref{fig:fig1} a) and b).

Mesons are, as usual, strings between a quark-antiquark pair.\\

\subsection{The chiral aspects of the Lagrangian}

In this subsection we briefly mention relevant aspects of chiral
symmetry, which we insist on having properly implemented. The
pairing force is chirally symmetric. On the other hand, following
the procedure of ref \cite{Uehara}, the string interaction, which
is not automatically chirally invariant, may be also reconciled
with chiral symmetry. In our approach, the Dirac see is naturally
responsible for  the dynamical breaking of  chiral symmetry,
contrary to an artificial device such a Mexican hat. Our
calculation shall be based on effective  Lagrangians, either of
the Nambu-Jona-Lasinio (NJL) type, or of sigma model
type. We argue that 
both models are equivalent as far as
the mean field description of bulk static properties of hadronic matter is concerned.
However, they lead, in the RPA approximation, to manifestly distinct dynamics,
in spite of being equally consistent with chiral symmetry,
so that it really matters which model one should consider if, beyond bulk properties, we
are interested on the surface properties or on the response of the system to an external probe.

\section{The effective action and the Dirac equations}

In this section, we describe the general structure of the
effective action which is relevant for the present development
and we try to motivate,  on the basis of
the Dirac equation of quark dynamics, the model which will be
considered in the next section.
The effective interaction is supposedly obtained by integrating over the 
gauge fields.

Below we
describe all the terms appearing in the effective Lagrangian, where
the potential terms  are coming out of a typical cluster expansion
in the quark fields. First comes the kinetic term
$i\bar\psi\gamma^\mu\partial_\mu\psi;$ next come all the 4-fermion
terms making up the 2-body force,

$$
\bar\psi\psi\bar\psi\psi+(\barpsi\vec\tau\psi)
\cdot(\barpsi\vec\tau\psi)-(\barpsi\gamma_5\psi)
\cdot(\barpsi\gamma_5\psi)-(\barpsi\vec\tau\gamma_5\psi)
\cdot(\barpsi\vec\tau\gamma_5\psi);\nonumber
$$
which, following a common procedure, we, for simplicity, replace by
the combination which is familiar from the NJL model \cite{Klevansky92},
\begin{equation}
\bar\psi\psi\bar\psi\psi
-(\barpsi\vec\tau\gamma_5\psi)
\cdot(\barpsi\vec\tau\gamma_5\psi).
\end{equation}
Finally, come the 6-quark terms contributing to the 3-body string
force. In order to insure  chiral invariance these three body
forces can be understood as reductions of appropriate chiral
invariant  4-body forces.

It is assumed that the
string force is responsible for the
transformation of the quark degrees of freedom into nucleon and
$\sigma,\,\vec\pi$ meson degrees of freedom, in the strong coupling limit, at low
energies.
The $\sigma$-nucleon
coupling may be taken three times as large as the $\sigma$-quark
coupling. However, as shown 
in Ref. \cite{0),01)}, the quark dynamics arising from the
confining force has the consequence that the nucleon coupling
constant decreases as the baryonic density increases. It will be argued that
this effect is important for saturation of nuclear matter.

In Ref. \cite{0)}
the confinement of quarks into nucleons is described on the basis
of the MIT bag model through the solution of an appropriate Dirac
equation,
\begin{equation}
i\partial_t\psi=-i\vecalpha\cdot\vecnabla\psi+V_{MIT}\psi+g\beta\sigma\psi.
\end{equation}
The $\sigma$ field is determined self-consistently, its source
being the scalar density due to $\psi.$ If the MIT interaction
$V_{MIT}$ \cite{2)} is replaced by a string force $m(r)=\kappa r$, where
$\kappa^2$ is the string tension, in a form which is consistent
with chiral symmetry, the previous equation may be replaced by the
following one \cite{Uehara}
\begin{equation}i\partial_t\psi=-i\vecalpha\cdot\vecnabla\psi+
\beta (f_{\pi}^{-1}m(r)+g)
(\sigma-i\gamma_5\vec{\tau}\cdot\vec\pi)\psi,
\end{equation}
where $\sigma,\,\vec\pi$ denote, respectively, the sigma meson and
the pion field, and
$f_{\pi}$ is the pion decay constant in the vacuum.
Here, we have already factored out the
three-body wave function into a product
described by three independent equations.
There are important differences between 
the models in Refs. \cite{0)} and \cite{Uehara} which
explain the
extra factors we introduce, {namely $(f_{\pi}^{-1}m(r)+g)$},
and  which are of interest for the present discussion,
especially due to the fact that  we insist on having chiral
invariance. Moreover, the $\sigma$ field in our case couples to
positive and negative energy states and therefore is nonzero in
the vacuum. The relevant consequence of this, which will be
explained later with respect to the phase analysis, is that we
expect a much stronger saturation effect at high density of
hadronic matter, and hence an enhancement of asymptotic freedom
phenomena for quarks.

\def\E{{\varepsilon}}
We shall now check the consistency of our string configuration and
the string potential $m(r)$ by analyzing the 
{spectrum of the Dirac operator:}
\begin{equation}
[\alpha_rp_r+{i\over r}\alpha_r\beta k+
\beta\sigma (f_{\pi}^{-1}m(r)+g)]\psi=\E(\sigma)\psi.
\end{equation}
where
\begin{equation}
\alpha_r=\left[\begin{array}{cc}0&-i\\i&0\end{array}\right]\quad
\quad
\beta=\left[\begin{array}{cc}1&0\\0&-1\end{array}\right]
\end{equation}
and $k=\pm1,\pm2,\cdots$.
Here we have used spherical symmetry. We take, for $r<1$fm,
$g+f_\pi^{-1}m(r)=\kappa f_\pi^{-1} r$, where $\kappa$
is the same as
defined in connection with eq. (\ref{2.1}). This gives us then,
\begin{equation}
\psi(r)={1\over
r}\left[\begin{array}{c}F(r)\\G(r)\end{array}\right]\end{equation}
\begin{eqnarray}(\E(\sigma)-\sigma f_\pi^{-1}\kappa r)F+{\d G\over \d r}
+{k\over r}G &=&0\nonumber\\
(\E(\sigma)+\sigma f_\pi^{-1} \kappa r)G-{\d F\over \d r}+{k\over
r}F &=&0\end{eqnarray} Eliminating one of the variables we obtain
a second order equation in the other which may be solved
numerically.
\def\kappad{{\sigma f_{\pi}^{-1}\kappa\over \varepsilon^2}}
The equations will have the form

\def\d{{\rm d}}\def\r{{\rho}}
\begin{equation}
\left({1}
+\kappad\r\right)G+\left({\d\over\d\r}-{k\over\r}\right)\left({{1} -\kappad\r}\right)^{-1}
\left({\d\over\d\r}+{k\over\r}\right)G=0\end{equation} and
similarly for $F$. We have defined
$ \r=\varepsilon r.$

A few qualitative observations can now be made on the basis of
this equation. This equation is singular for
$\sigma f_{\pi}^{-1}\kappa=\pm\varepsilon^2$.This gives
certain rules for the energy eigenvalue $\E(\sigma)$. It should be noticed
that the string interaction is proportional to the $\sigma$ field.

The QCD vacuum, 
which is studied in these calculations, is a very complex object
not least because of the strong contributions from both positive
and negative energy states. In our model it is described as a
collection of nucleonic bags arising out of the solution of the
Dirac equation, in all possible momentum states ${\bf P}$ such
that $|{\bf P}|\leq \Lambda$, where $\Lambda$ is a regularizing
cut off. In the framework of the independent particle approach
which has been described, the mass of the nucleonic bag is
approximately $M(\sigma)=3\E(\sigma)$. The scalar density $\rho_S$
and the energy of the vacuum are also important quantities and
their calculation, which is left to another section, is very
informative concerning this point.
The vacuum 
value of the $\sigma$ field is set to
$f_{\pi}$, i.e. $ \sigma  = f_{\pi}$ if the Fermi momentum vanishes, $p_F=0$.

The asymptotic solution for $G$ and $F$ fields in the Dirac
equation can easily be found by approximating the equation to
$$G-{\varepsilon^2\over \kappa\r}{\d\over\d \r}\left(
{\varepsilon^2\over \kappa\r}{\d\over\d \r}G\right)=0.$$ Thus the asymptotic
solution for $G$ and $F$ will be
$$G\sim\exp(-{\kappa\r^2\over 2\varepsilon^2}).$$
Here we can easily read out the size of the nucleon bag to be
$1/\sqrt{\kappa}$, which becomes approximately 0.8 fm, in vacuum.
\section{The model}
{To be more specific, we present now in greater detail the
basic scheme which supports our calculation.  This
framework is motivated by the previous discussion and is based on
the assumption that quarks and gluons are the building blocks of
nucleons. We assume that the interaction between quarks originates
from an instanton force and from a confining string interaction,
the latter being responsible for the clustering of quarks into
nucleons and the former for the emergence of a sigma field which
mediates the interaction between nucleons and is understood as a
quark-antiquark disturbance of the vacuum. According to this picture,
the well known sigma model Lagrangian density may be extended to read
\begin{eqnarray}
&&{\cal
L}=i\bar\psi\gamma^\mu\partial_\mu\psi-g_s(\sigma,\vec\pi)[(\bar\psi\psi)\sigma
+i(\bar\psi\gamma^5\vec\tau\psi)\cdot\vec\pi]+g_v(\bar\psi\gamma_\mu\psi)\omega^\mu\nonumber\\
&&-
{1\over2}m_\sigma^2(\sigma^2+\vec\pi^2)+{1\over2}m_v^2\omega_\mu\omega^\mu
+{1\over2}(\partial^\mu\sigma\partial_\mu\sigma+
\partial^\mu\vec\pi\cdot\partial_\mu\vec\pi)-{1\over4}\omega^{\mu\nu}\omega_{\mu\nu},
\label{linearsigma}
\end{eqnarray}
where $\psi$ denotes the nucleon field (not the quark field). This Lagrangian density
describes an assembly of nucleons, regarded as composite particles, which interact with
$\sigma,\vec \pi$ and $\omega^\mu$ fields.}
The familiar Mexican hat does not appear in eq. (\ref{linearsigma}). However,
terms involving the $\omega$ field are introduced, as in the Walecka model \cite{walecka}.
In a conventional model, with elementary
particles, $g_s$ is a constant. In our model,
which we call extended sigma model (ESM), the composite nature of the particles
reflects itself in the fact that
$g_s$
depends on the local values of the fields $\vec\pi, \sigma$. We assume that
\begin{equation}
g_s(\sigma,\vec\pi)=g_0\sqrt{\frac{1-\sqrt{1-4g_0^2a^2(\sigma^2+\vec\pi^2)}}{2g_0^2a^2(\sigma^2+\vec\pi^2)}}\approx g_0(1+{1\over2}g_0^2a^2(\sigma^2+\vec\pi^2))
\label{g_s}
\end{equation}
where $a, g_0$ are phenomenological parameters. For an extended
system, $\vec\pi=0.$ The dynamically generated mass of the nucleon
becomes then $M(\sigma)=g_s(\sigma)\sigma$. From eq. (\ref{g_s})
it follows that
\begin{equation}M(\sigma)=g_0\sigma\sqrt{\frac{1-\sqrt{1-4g_0^2a^2\sigma^2}}{2g_0^2a^2\sigma^2}}
\approx g_0\sigma(1+{1\over2}g_0^2a^2\sigma^2).
 \label{M(sigma)}\end{equation}

We have circumvented the laborious problem of determining the mass
$M(\sigma)$ of a nucleon subject to an external field $\sigma$ by
making an ansatz. According to our assumption,  $M(\sigma)$
behaves as $g_0\,\sigma$ for small $\sigma,$ but increases faster
for large $\sigma$.
We also have
$$\sigma^2={M^2\over g_0^2}(1-a^2M^2).
$$
The energy of an assembly of 
nucleons is
\begin{eqnarray}
{\cal E}(\sigma,\omega)&=& -\eta\sum_{p_F\leq|{\bf P}|\leq
\Lambda}\sqrt{{\bf P}^2+g_s^2\,\sigma^2} -\eta\sum_{|{\bf P}|\leq
p_F}g_v\omega
+{1\over2}m_\sigma^2\sigma^2V-{1\over2}m_v^2\omega^2V,\nonumber
\end{eqnarray}
or, after {eliminating $\sigma$ in favor of $M$ and}
``minimization" w.r.t. $\omega,$
\begin{eqnarray}
{\cal E}(\sigma)&=& - \eta\sum_{p_F\leq|{\bf P}|\leq
\Lambda}\sqrt{{\bf P}^2+M^2} +{m_\sigma^2\over
2g_0^2}M^2(1-a^2M^2) V+{g_v^2p_F^6\eta^2\over 2^33^2 m_v^2\pi^4 }V
\label{E(sigma)}.
\end{eqnarray}
Here, $V$ is the normalization volume, $p_F$ is the Fermi
momentum, $\Lambda$ is the regularizing cut-off momentum and
$\eta=2N_f$ is the degeneracy. {The natural assumption has been
made that the wave function of the assembly of nucleons is a
Slater determinant}. Minimizing ${\cal E}(\sigma)$, we obtain the
generalized gap equation
\begin{equation}
{g_0^2\eta\over m_\sigma^2V}
\sum_{p_F\leq|{\bf P}|\leq
\Lambda}{1\over\sqrt{{\bf P}^2+M^2}}
 =
1-2a^2M^2.\label{gengap}
\end{equation}

The coupling constant
 depends on $\sigma$, and,
through $\sigma$, on the baryonic density.  The coupling constant
increases essentially linearly with $\sigma^2$. The practical
consequence of such a behavior is the emergence of a rapid
increase of the energy with density, tantamount to the onset of a
repulsive contribution. This behavior plays a decisive role in
fixing the saturation density.
\section{NJL versus sigma model}
In this section we discuss briefly the equivalence of the sigma model to the
Nambu-Jona-Lasinio (NJL) model
 \cite{Klevansky92},
as far as the mean field description
of bulk properties of hadronic matter is concerned.
The NJL model
is defined by the Lagrangian density
\begin{equation}{\cal L}=\bar\psi(i\gamma^\mu\partial_\mu)\psi+{G_S\over2}[(\bar\psi\psi)^2
+(\bar\psi i\gamma_5\vec\tau\psi)^2]. \label{Lag2}
\end{equation}
The sigma model is defined by
 eq. (\ref{linearsigma}),
provided we replace
the coupling parameter $g_s(\sigma,\vec\pi)$ by a constant.
 A regularizing momentum cut-off $\Lambda$ is part of 
both models.
 Using standard procedures,
the Lagrangian (\ref{Lag2}) leads to the Hamiltonian
\begin{equation}
{\cal H}_{NJL}=\sum_{k=1}^N\vecp_k\cdot\vecalpha_k+
{G_S\over2}\sum_{k,l=1}^N\delta(\vecr_k-\vecr_l)
\beta_k\beta_l(1-\gamma^5_k\gamma^5_l\vectau_k\cdot\vectau_l).
\end{equation}
The vacuum is described by a Slater determinant, $|\Phi_0\rangle$,
created by the operators $b_{\vecp,-}^{\dag},$
satisfying $|\vecp|<\Lambda$, associated with
plane wave negative energy
eigenstates of the single particle Hamiltonian
$h=\vecp\cdot\vecalpha+\beta M,$
and by the operators $b_{\vecp,+}^{\dag}$,
satisfying $|\vecp|<p_F$, so that $p_F$ is the
Fermi momentum,
associated with positive energy eigenfunctions
of the same operator.
 The {nucleon} ``constituent mass'' $M$ is a variational parameter.


\def\E{\varepsilon}
The energy expectation value
${\cal E}=\langle \Phi_0|{\cal H}_{NJL}|\Phi_0\rangle$ reads
\begin{eqnarray}
{\cal E}
&=&-\eta\sum_{|\vecp|=p_F}^\Lambda\frac{P^2}{\E_P}-\frac{G_S}{2V}
\left[\eta\sum_{|\vecp|=p_F}^\Lambda\frac{M}{\E_P}\right]^2\nonumber\\
&=&-\frac{\eta V}{2\pi^2}\int_{p_F}^\Lambda{\rm d}P\frac{P^4}{\E_P}
-\frac{G_SV}{2}\left[\frac{\eta}{2\pi^2}\int_{p_F}^\Lambda{\rm d}P
\frac{MP^2}{\E_P}\right]^2\label{(23)}
\end{eqnarray}
where $\E_P=\sqrt{P^2+M^2}$, $V$ is the normalization volume,
$\eta$ is the degeneracy and, for nuclear matter, $G_S$ is 9 times
bigger than the corresponding quark matter value. For quark
matter, $\eta=2N_cN_f$, while for nuclear matter, $\eta=2N_f$. The
condition $\partial{\cal E}/\partial M$=0 leads to the gap
equation
\begin{equation}
1=\frac{\eta G_S}{V}\sum_{|\vecp|=p_F}^\Lambda\frac{1}{\E_P}=
\frac{\eta G_S}{2\pi^2}\int_{p_F}^\Lambda{\rm d}P\frac{P^2}{\E_P},
\end{equation}
which fixes $M$. For nuclear 
matter we fix $\Lambda$ so that $M=939$
MeV. For quark matter we  fix $\Lambda$ so that 
$M=313$ MeV. If we take
$$G_S={9g_0^2\over m_\sigma^2},$$
 this is essentially the same gap equation as (\ref{gengap}),
except that the quantity $1-2a^2M^2$ in the rhs of (\ref{gengap}) is here replaced by 1.

Notice also that the minimum of ${\cal E}$ given by
eq.(\ref{(23)}) is the same as the minimum of ${\cal E}(\sigma)$
given by eq.(\ref{E(sigma)}), if $a$ is set equal to 0 in the
expression for $M(\sigma)$, eq.(\ref{M(sigma)}).

{An extended NJL (ENJL) model equivalent to (\ref{linearsigma})
is easily obtained and reads
\begin{eqnarray}&&{\cal L}_{ENJL}=\bar\psi(i\gamma^\mu\partial_\mu)\psi+{G_S\over2}[(\bar\psi\psi)^2
+(\bar\psi i\gamma_5\vec\tau\psi)^2]\nonumber\\&&+{K\over12}[(\bar\psi\psi)^2
+(\bar\psi i\gamma_5\vec\tau\psi)^2]^2+{G_V\over2}[(\bar\psi\gamma_\mu\psi)(\bar\psi\gamma^\mu\psi)
]. \label{ENJL}
\end{eqnarray}
A similar model has been studied in Ref. \cite{buervenich}.

{\footnotesize
\begin{table}[!ht]
\caption{\small{Parameters of ESMI and ESMII } }
\begin{center}
\begin{tabular}{lcccc}
\hline
&$g_s$&$g_v$& $\Lambda$ (MeV)&$a$\\
\hline
ESM I &5.81 & 10.75& 477 & 0.30\\
ESM II &8.12& 13.16&387 &0.25\\
\hline
\end{tabular}
\end{center}
\label{tableI}
\end{table}}

{\footnotesize
\begin{table}[!ht]
\caption{\small{Properties of nuclear matter
according to ESMI and ESMII: binding energy, equilibrium density,
incompressibility, $f_\pi$, chiral transition density and deconfinement density. } }
\begin{center}
\begin{tabular}{lccccccc}
\hline
&$E_B$& $\rho_0$& $M/M_0$ & $K$ &$f_\pi$ &$\rho_\chi$&$\rho_{de}$\\
&(MeV)& (fm$^{-3}$)& & (MeV)& (MeV)&(fm$^{-3}$)&(fm$^{-3}$)\\
\hline
ESM I & -15.75& 0.15&0.7&225 &150&0.46&0.687\\
ESMII & -15.75 &0.148&0.58& 240&108&0.3&0.36\\
\hline
\end{tabular}
\end{center}
\label{tableII}
\end{table}}
\bigskip
\section{The phases of nuclear matter and quark clustering}

In order to get a good understanding of the various phases nuclear matter
can exhibit at different energies, it is useful to introduce some descriptors
for the various nuclear processes. We usually describe quantum field
interactions in terms of the momentum transfer of the strong interaction
scattering involving quarks and gluons. One can as well picture the
energy per nucleon as a function of pressure.

There are two distinct transitions that are important in the
nuclear phase diagram \cite{Luso} when varying pressure or
momentum transfer. One is the chiral restoration
transition, mostly determined by the instanton pairing force
\cite{hooft,1)}. Secondly there is the
quark deconfinement transition when
the nuclear ``fluid"  changes from ``water'' to becoming
``metallic fluid''. This is a consequence of the diminishing
baryonic string force,
predominantly due to the behaviour of the
effective coupling constant.

We wish now to calculate the energy per particle of nuclear matter
as a function of the density.
As pointed out before, our two main ingredients are the pairing
force and the baryonic string force. We take these contributions separately
since they apply mostly to different scales.
The effect of the string force is to confine quarks within nucleons,
while the pairing force comes from  instantons.
\begin{center}
\begin{figure}[t]
\begin{center}
\epsfig{file=
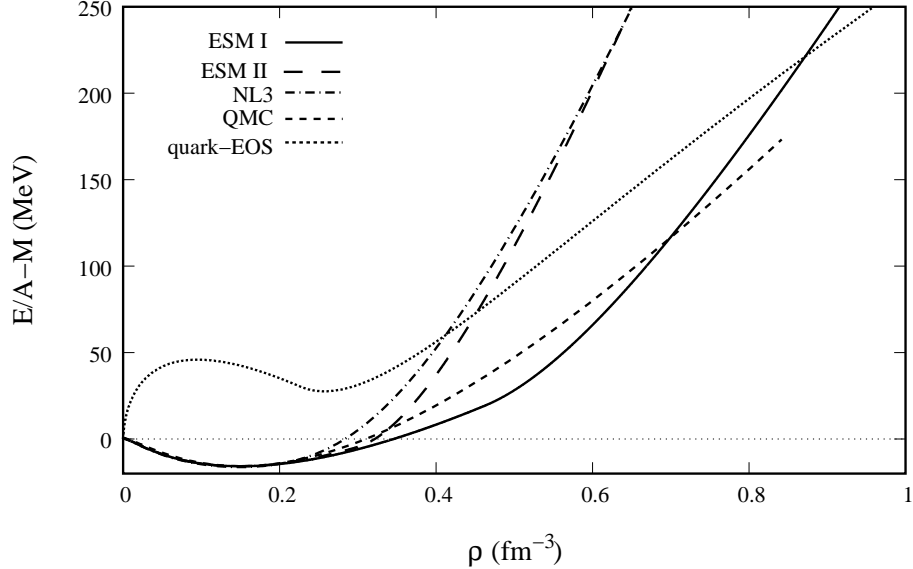,width=12cm } 
\caption{Phase diagram of  hadronic matter energy as a function of the density.
The
dotted line applies to a quark gas. The dash-dotted curve
corresponds to the
so-called NL3 parametrization of the Walecka model \cite{ring}.
The full curve and the long dashed curve present, respectively,
the results of ESMI and ESMII.
The dashed curve corresponds to the
quark-meson coupling model \cite{0)}. \label{fig:fig5}}
\end{center}
\end{figure}
\end{center}

We have considered two extreme parametrizations of the ESM,
denoted ESMI and ESMII. The  corresponding results are shown in
the tables and figures. The parameters of the model, $g_s,\,g_v$
and $a$, were chosen in order to reproduce the saturation
properties of nuclear matter, namely, binding energy, saturation
density and effective mass at saturation. The compressibility and
$f_\pi$ are outputs of the calculation. We are now able to make
the plots of the energy per nucleon as a function of density,
shown in Fig. \ref{fig:fig5}. The dotted line applies to a quark
gas described by the Lagrangian (\ref{Lag2}) with quarks. The
coupling constant $G_S=2.14/\Lambda^2$ was chosen so  that
$f_\pi=93$ MeV and $\Lambda=654$ MeV. This cut-off reproduces the
correct quark mass in vacuum, $m_q=$313 MeV.
  We see that the quark equation of state exhibits no binding,
although it shows a local minimum
corresponding to chiral symmetry restoration.
This
occurs at around 0.2 fm$^{-3}$
which corresponds to the generally accepted deconfinement
transition expected at a temperature of about 200 MeV in the
vacuum.
The full curve and the long-dashed curve represent the effect of quark clustering
according to ESMI and ESMII, respectively.
The cutoff $\Lambda$ is such that chiral
symmetry breaking of the vacuum reproduces the correct nucleon mass. The clear
binding shown by
these curves is due to the reduction in
the kinetic energy arising from the clustering of 3 quarks into
nucleons  \cite{Moszk}.
This latter behavior is crucial for stability.
Our model, not only takes into account the effect of quark clustering,
but  also the repulsion between the nucleonic bags. {To account for
the interplay between these effects we have used eqs.(\ref{linearsigma},
\ref{g_s}, \ref{E(sigma)}, \ref{gengap}).
In Fig. \ref{fig:fig5}, chiral symmetry restoration is shown as a discontinuity of
the second derivative of the ESMI and ESMII curves. Deconfinement occurs
when these curves intercept the quark-EOS dotted curve.
Although our calculation is indeed based on the ESM (eq.(\ref{linearsigma})), we observe
that it would equally be possible to use
the ENJL model (eq.(\ref{ENJL})), which also
reflects the influence of the medium on the nucleon properties
(see, for instance, ref. \cite{buervenich} or ref. \cite{Koch87},
where a density dependent coupling is used,
in contrast to our assumption of a
coupling depending on the local value of the sigma field).}
However, we insist that the equivalence between the ESM
(eqs. (\ref{linearsigma}, \ref{g_s})) and the ENJL model
(eq.(\ref{ENJL})) applies only to the description of bulk
properties. These models differ in the description of surface
properties and in the predicted dynamics, such as the masses of
scalar excitations, for which they lead to distinct dispersion
relations.}

As has been argued in Ref. \cite{0)}, the quark structure of the nucleon induces a
mechanism for saturation by weakening, at high density, the attraction due to the
sigma meson. We find that the contribution of this effect
to the energy per particle is equivalent to a term essentially
proportional to $\rho^2.$
In order to fit the high energy
scattering nuclear experiment it is important that the slope in
the high energy end (large pressure) is not too steep meaning a
fluid of nuclear matter that is not too incompressible.
\begin{center}
\begin{figure}[t]
\begin{center}
\epsfig{file=
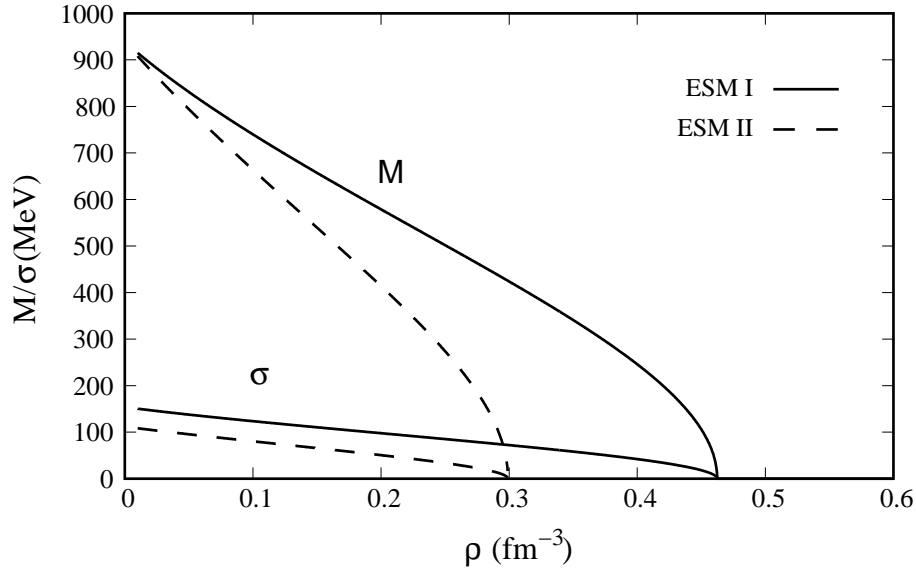,width=12cm }
\caption{Effective mass of the nucleon and sigma field as function of
 the density for ESMI and ESMII.
Chiral symmetry restoration occurs when these quantities vanish.
\label{fig:fig6}}
\end{center}
\end{figure}
\end{center}

\begin{center}
\begin{figure}[t]
\begin{center}
\epsfig{file=
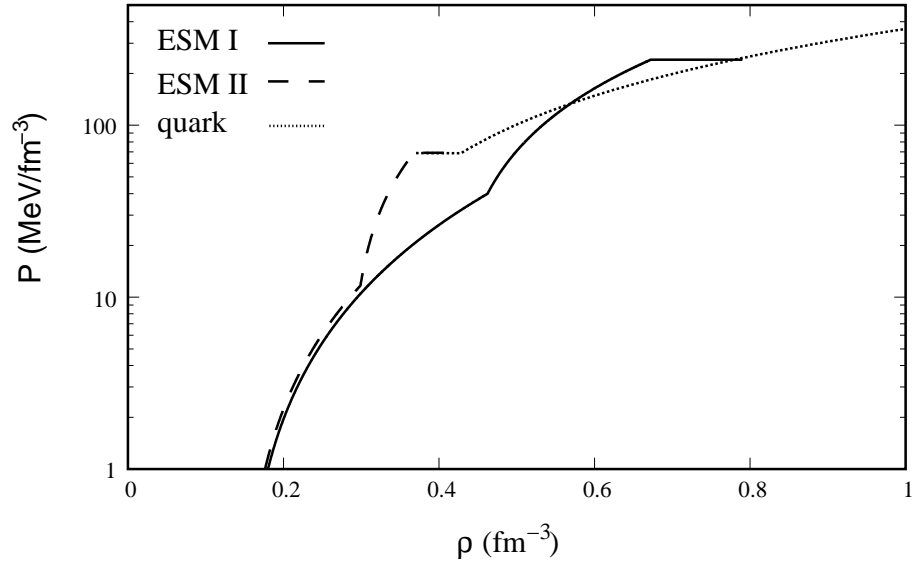,width=12cm } 
\caption{Pressure versus density for ESMI and ESMII. Phase transitions
are shown as discontinuities in the slopes of the curves.
 \label{fig:fig7}}
\end{center}
\end{figure}
\end{center}

In principle, one may expect to find
two critical limits: the limit of chiral symmetry restoration,
when the mass of the nucleon vanishes, and the deconfinement
limit, when the baryonic string force becomes small and quarks get
liberated. For ESMI,
chiral symmetry restoration
is depicted in Figs. \ref{fig:fig5}, 
\ref{fig:fig6} and \ref{fig:fig7}
 at $\rho/\rho_0\approx 3.1,$  
while the transition from the confined phase to a deconfined phase
is shown in Figs \ref{fig:fig5}  and \ref{fig:fig7} at
$\rho/\rho_0\approx 4.5$.
For ESMII,
chiral symmetry restoration
is depicted at $\rho/\rho_0\approx 2.0,$
and the transition from the confined phase to a deconfined
phase takes place at $\rho/\rho_0\approx 2.4$.
We remark that the equation
of state predicted by
ESMI is consistent with the one
predicted by the quark-meson coupling model \cite{0)}, and is much
softer than the one predicted by the so called NL3 parametrization
of the Walecka model \cite{ring}.
The equation of state predicted by
ESMII is also rather soft. Our model is consistent with the
occurrence of a superfluid exotic phase of hadronic matter between
the critical points corresponding to chiral symmetry restoration
and deconfinement. In particular, this phase might be composed by
Bose particles, each particle being constituted by six quarks.
Our model is also consistent with the
occurrence of the so called CFL phase, above the second critical
point.

It may also be observed that the values of $f_\pi$ predicted by
the ESMI and by the ESMII and shown in Table \ref{tableII},
respectively 150 and 108, are quite reasonable.

Finally, we wish to mention a recent attempt pursued by Chanfray and
collaborators \cite{chanfray}
to reconcile the Walecka model with chiral symmetry, although along different lines.

\section{Summary}
In the present article we have constructed
a realistic quark model of nuclear structure phenomena which is
inspired on the Quantum Chromo Dynamics (QCD) quantum field theory,
being formulated with very few free parameters, {namely $\Lambda,a,g_0,g_v$}.
In its final form, our model resembles the conventional NJL model for
nucleons with, however, consequential differences reflecting the influence
of the medium on the properties of the nucleons.
In order to meet important requirements of chiral symmetry arising
in the  description of nuclear phenomena at a sub-nuclear level,
we have added to the starting Lagrangian which describes
the dynamics at the quark scale,
important terms
coming from chiral fields
that appear at a higher
order of QCD, in a cluster expansion sense. Such a unified
model of nuclear physics 
is consistent with important data about nuclear phases from low
energy physics and astrophysics \cite{Glend}, such as chiral
symmetry restoration and quark deconfinement.\\


\section*{Acknowledgment}


We wish to thank Prof. K. Yazaki and Prof. J. Clark for helpful
comments. One of us, (J.P.), wishes to thank S. Moszkowski, H.
Walliser, Y. Tsue and P. V. Alberto for valuable discussions.

This work was supported by the  Foundation for Science and Technology (Portugal)
through the the Project POCTI/ FIS/451/94.

\subsection*{Appendix: RPA treatment of the NJL and extened sigma models}

\def\d{{\rm d}}
\def\vecalpha{{\mbox{\boldmath{$\alpha$}}}}
\def\vecnabla{{\mbox{\boldmath{$\nabla$}}}}
We argue that, although the extended NJL model and the extended sigma model proposed here
are equivalent as far as
the mean field description of bulk static properties of hadronic matter is concerned,
they lead, in the RPA approximation, to manifestly distinct dynamics.

For simplicity, we forget about
the $\sigma$ dependence of the coupling coeficient $g_\sigma$.
The Lagrangian (\ref{linearsigma}) is chiral invariant.
Therefore, one of the RPA modes comes at zero energy. This is the
pion. Avoiding straightforward algebraic developments,
we present the quadratic hamiltonian which describes the RPA modes
having 0 momentum,
\def\veck{{\bf k}}
\begin{eqnarray}
&{\cal H}^{(2)}&=\sum_{p_F<|\veck|<\Lambda} \varepsilon_k
(c^{\dag}_{\veck,\tau} c_{\veck,\tau}+d^{\dag}_{\veck,\tau}
d_{\veck,\tau})+m_\sigma
(a^{\dag}_0a_0+b^{{0\dag}}_0b^{0}_0+b^{+{\dag}}_0b^{+}_0+b^{-{\dag}}_0b^{-}_0)\nonumber
\\&&+{g\over\sqrt{2 m_\sigma V}}\sum_{p_F<|\veck|<\Lambda}
{|\veck|\over\varepsilon_k}
(c^{\dag}_{\veck,\tau}
d^{\dag}_{-\veck,-\tau} -c_{\veck,\tau}d_{-\veck,-\tau})(a_0+a_0^{\dag})\nonumber\\
&&+{g\over\sqrt{2 m_\sigma V}}\sum_{p_F<|\veck|<\Lambda}
\left[c^{\dag}_{\veck,\tau}
d^{\dag}_{-\veck,-\tau'}\left({T^{+}_{\tau,\tau'}}
(b^{-{\dag}}_0+b^{+}_0)+{T^{-}_{\tau,\tau'}}
(b^{+{\dag}}_0+b^{-}_0)+{2T^{0}_{\tau,\tau'}}
(b^{0{\dag}}_0+b^{0}_0)\right)\right.\nonumber\\
&&\left.-c_{\veck,\tau} d_{-\veck,-\tau'}
\left({T^{+}_{\tau,\tau'}}(b^{+{\dag}}_0+b^{-}_0)+{T^{-}_{\tau,\tau'}}
(b^{-{\dag}}_0+b^{+}_0)+{2T^{0}_{\tau,\tau'}}(b^{0{\dag}}_0+b^{0}_0)\right)\right]\,.\nonumber
\end{eqnarray}
Here, $\varepsilon_k=\sqrt{\veck^2+M^2},$ $c^{\dag},c,d^{\dag},d$ are
fermion and anti-fermion operators, $a^{\dag},a$ are scalar meson
operators and
$b^{{0\dag}},b^{0},b^{+{\dag}},b^{+},b^{-{\dag}},b^{-}$ are
pseudo-scalar meson operators. The quantities
$(T^{+}_{\tau,\tau'}+T^{-}_{\tau,\tau'}),\,i(T^{+}_{\tau,\tau'}
-T^{-}_{\tau,\tau'}),\,2T^{0}_{\tau,\tau'}$ are
Pauli matrices, or, more precisely, they are tensor products of
Pauli matrices in isospin times identity matrices in spin, the
indices $\tau$ being pairs of spin-isospin indices. Summation over
repeated $\tau$, $\tau'$ indices is understood.
Terms responsible for the excitation of finite momentum RPA modes
have been deliberately omitted. The quantities
$$c^{\dag}_{\veck,\tau}d^{\dag}_{-\veck,-\tau},\quad c^{\dag}_{\veck,\tau}d^{\dag}_{-\veck,-\tau'}T^{j}_{\tau,\tau'},\quad
d_{-\veck,-\tau'}c_{\veck,\tau}T^{j}_{\tau,\tau'},\quad j\in\{+,-,0\}
$$
behave as quasi-bosons. For instance
$$[d_{-\veck,-\tau'''}c_{\veck,\tau''}
,c^{\dag}_{\veck,\tau}d^{\dag}_{-\veck,-\tau'}]\,
T^{j}_{\tau'',\tau'''}T^{k}_{\tau,\tau'}
\approx\delta_{jk}\eta_k,
\quad\eta_+=\eta_-=4,\quad\eta_0=2,
$$
$$[d_{-\veck,-\tau'}c_{\veck,\tau'}
,c^{\dag}_{\veck,\tau}d^{\dag}_{-\veck,-\tau}]
\approx 4.
$$
The result requires replacing such quantities as  $d^{\dag}_{-\veck,-\tau'}d_{-\veck,-\tau'''}$ and
$c^{\dag}_{\veck,\tau}c_{\veck,\tau''}$ by 0, which is their ground state expectation value.
Notice also that
$$
[c_{\veck,\tau''}^{\dag}c_{\veck,\tau''},
c^{\dag}_{\veck,\tau}d^{\dag}_{-\veck,-\tau'}]\,T^{k}_{\tau,\tau'}=
[d_{-\vec k,-\tau''}^{\dag}d_{-\vec k,-\tau''},
c^{\dag}_{\veck,\tau}d^{\dag}_{-\veck,-\tau'}]\,T^{k}_{\tau,\tau'}=
c^{\dag}_{\veck,\tau}d^{\dag}_{-\veck,-\tau'}T^{k}_{\tau,\tau'},
$$
As an example,
we present now more explicitly the computation of charged pseudoscalar modes.
The following commutators are easily obtained
\begin{eqnarray}
&[{\cal H}^{(2)},c^{\dag}_{\veck,\tau}d^{\dag}_{-\veck,-\tau'}T^{+}_{\tau,\tau'}]
&\approx 2\varepsilon_k
c^{\dag}_{\veck,\tau}d^{\dag}_{-\veck,-\tau'}
T^{+}_{\tau,\tau'}+{4g\over\sqrt{2 m_\sigma V}}(b_0^-+b_0^{+\dag})\,,\nonumber
\\
&[{\cal H}^{(2)},c_{\veck,\tau}d_{-\vec k,-\tau'}T^{-}_{\tau,\tau'}]
&\approx -2\varepsilon_k
c_{\vec k,\tau}d_{-\veck,-\tau'}T^{-}_{\tau,\tau'}+{4g\over\sqrt{2 m_\sigma V}}(b_0^-+b_0^{+\dag})\,,\nonumber\\
&\quad\quad\quad\quad[{\cal H}^{(2)},b^{+\dag}_0]&=m_\sigma b_0^{+\dag}+{g\over
\sqrt{2 m_\sigma V}}\sum_{p_F<|\vec k|<\Lambda}
(c^{\dag}_{\veck,\tau}d^{\dag}_{-\veck,-\tau'}T^{+}_{\tau,\tau'}
-c_{\veck,\tau} d_{-\veck',-\tau}T^{-}_{\tau,\tau'})\,,\nonumber\\
&\quad\quad\quad\quad[{\cal H}^{(2)},b_0^-]&=-m_\sigma b_0^--{g\over\sqrt{2 m_\sigma
V}} \sum_{p_F<|\veck|<\Lambda}
(c^{\dag}_{\veck,\tau}d^{\dag}_{-\veck,-\tau'}
T^{+}_{\tau,\tau'}-c_{\veck,\tau} d_{-\veck,-\tau'}T^{-}_{\tau,\tau'})\,.
\nonumber\end{eqnarray}
It follows that
\begin{eqnarray}&&\left[{\cal H}^{(2)},\sum_{p_F<|\veck|<\Lambda}
x_{\veck}(c^{\dag}_{\veck,\tau}d^{\dag}_{-\veck,-\tau'}T^{+}_{\tau,\tau'}
+c_{\veck,\tau}d_{-\veck,-\tau'}T^{-}_{\tau,\tau'})-(b_0^--b_0^{+\dag})\right]\nonumber\\
&&\approx\sum_{p_F<|\veck|<\Lambda}\left(2x_{\veck}\varepsilon_k+{2g\over
\sqrt{2 m_\sigma V}}\right)
(c^{\dag}_{\veck,\tau}d^{\dag}_{-\veck,-\tau'}T^{+}_{\tau,\tau'}-
c_{\veck,\tau}d_{-\veck,-\tau'}T^{-}_{\tau,\tau'})\nonumber\\
&&+\left(\sum_{p_F<|\veck|<\Lambda}{8gx_{\veck}\over\sqrt{2 m_\sigma
V}}+m_\sigma\right) (b_0^-+b_0^{+\dag})=0.\nonumber
\end{eqnarray}
The result clearly follows from the gap equation, for
$$x_{\veck}=-{g\over\sqrt{2m_\sigma V}\,\varepsilon_k}\,.
$$
This is the RPA pseudo-scalar mode and its eigenfrequency is 0.

Similarly, we find that the eigenfrequency of  the RPA scalar mode
is close to $m_\sigma.$
We present  the computation of scalar modes.
The following commutators are easily obtained
\begin{eqnarray}
&[{\cal H}^{(2)},c^{\dag}_{\veck,\tau}d^{\dag}_{-\veck,-\tau}]
&\approx 2\varepsilon_k
c^{\dag}_{\veck,\tau}d^{\dag}_{-\veck,-\tau}
+{4g\over\sqrt{2 m_\sigma V}}{|\veck|\over\varepsilon_k}
(a_0^-+a_0^{+\dag})\,,\nonumber
\\
&[{\cal H}^{(2)},c_{\veck,\tau}d_{-\veck,-\tau}]
&\approx -2\varepsilon_k
c_{\veck,\tau}d_{-\veck,-\tau}
+{4g\over\sqrt{2 m_\sigma V}}{|\veck|\over\varepsilon_k}
(a_0^-+a_0^{+\dag})\,,\nonumber\\
&\quad\quad\quad\quad[{\cal H}^{(2)},a^{+\dag}_0]&=m_\sigma a_0^{+\dag}+{g\over
\sqrt{2 m_\sigma V}}\sum_{p_F<|\veck|<\Lambda}{|\veck|\over\varepsilon_k}
(c^{\dag}_{\veck,\tau}d^{\dag}_{-\veck,-\tau}
-c_{\veck,\tau} d_{-\veck,-\tau})\,,\nonumber\\
&\quad\quad\quad\quad[{\cal H}^{(2)},a_0^-]&=-m_\sigma a_0^--{g\over\sqrt{2 m_\sigma
V}} \sum_{p_F<|\veck|<\Lambda}{|\veck|\over\varepsilon_k}
(c^{\dag}_{\veck,\tau}d^{\dag}_{-\veck,-\tau}
-c_{\veck,\tau} d_{-\veck,-\tau})\,.
\nonumber\end{eqnarray}
It follows that
\begin{eqnarray}&&\left[{\cal H}^{(2)},\sum_{p_F<|\veck|<\Lambda}
\left(X_{\veck}c^{\dag}_{\veck,\tau}d^{\dag}_{-\veck,-\tau}
+Y_{\veck} c_{\veck,\tau}d_{-\veck,-\tau}\right)
-\eta a_0^-+\xi a_0^{+\dag}\right]\nonumber\\
&&\approx\sum_{p_F<|\veck|<\Lambda}2\varepsilon_k
(X_{\vec k}c^{\dag}_{\veck,\tau}d^{\dag}_{-\vec k,-\tau}
-Y_{\vec k}c_{\veck,\tau}d_{-\veck,-\tau}
)\nonumber\\
&&+\sum_{p_F<|\veck|<\Lambda}
{(\eta+\xi)g\over
\sqrt{2 m_\sigma V}}{|\veck|\over\varepsilon_k}
(c^{\dag}_{\veck,\tau}d^{\dag}_{-\veck,-\tau}
-c_{\veck,\tau}d_{-\veck,-\tau}
)\nonumber\\
&&+\sum_{p_F<|\veck|<\Lambda}{4g(X_{\veck}+Y_{\veck})\over\sqrt{2 m_\sigma
V}} {|\veck|\over\varepsilon_k}(a_0^-+a_0^{+\dag})\nonumber+
m_\sigma (\eta a_0^-+\xi a_0^{+\dag})\nonumber\\
&&=\Omega_s\left(\sum_{p_F<|\veck|<\Lambda}
\left(X_{\veck}c^{\dag}_{\veck,\tau}d^{\dag}_{-\veck,-\tau}
+Y_{\veck} c_{\veck,\tau}d_{-\veck,-\tau}\right)
-\eta a_0^-+\xi a_0^{+\dag}\right)\,.\nonumber
\end{eqnarray}
The RPA equations read
\begin{eqnarray}&&2\varepsilon_k
X_{\veck}+{(\eta+\xi)g\over
\sqrt{2 m_\sigma V}}{|\veck|\over\varepsilon_k}=\Omega_s
X_{\veck}\nonumber
\\
&&2\varepsilon_k
Y_{\veck}+{(\eta+\xi)g\over
\sqrt{2 m_\sigma V}}{|\veck|\over\varepsilon_k}=-\Omega_sY_{\veck}\nonumber
\\
&&\sum_{p_F<|\veck|<\Lambda}{4g(X_{\veck}+Y_{\veck})\over\sqrt{2 m_\sigma
V}} {|\veck|\over\varepsilon_k}+m_\sigma\xi=\Omega_s\xi\nonumber\\
&&\sum_{p_F<|\veck|<\Lambda}{4g(X_{\veck}+Y_{\veck})\over\sqrt{2
m_\sigma V}}
{|\veck|\over\varepsilon_k}+m_\sigma\eta=-\Omega_s\eta\nonumber
\end{eqnarray}
where $X_{\veck},Y_{\veck},\xi,\eta$ are the RPA amplitudes and $\Omega_s$
the renormalized sigma mass.

Let ${\cal G}_k={2g\over\sqrt{2 m_\sigma V}}{|\veck|
\over\varepsilon_k}.$ The following dispersion relation is
obtained
$$1={2m_\sigma\over\Omega^2-m_\sigma^2}\sum_{p_F<|\veck|<\Lambda}
{\cal G}_k^2{4\varepsilon_k\over\Omega^2-4\varepsilon_k^2}
$$
showing that $\Omega_s\approx m_\sigma<2M$. Indeed, the l.h.s. of
the dispersion equation becomes infinite for $\Omega^2=m_\sigma^2$
and for $\Omega^2=4\varepsilon_k^2,$ changing sign at each place.
The zero $\Omega_s^2$ lies 
slightly below $m_\sigma^2.$ 
Between $m_\sigma^2$ and $4M^2,$  the l.h.s. of the dispersion
equation is negative. Between $4M^2+4p_F^2,$  and $4M^2+\Lambda^2$ we
find a continuum of Landau damped modes.

Next we consider the quadratic hamiltonian which, for the NJL model, describes
the RPA modes having 0 momentum, 
\begin{eqnarray}
&{\cal H}^{(2)}_{NJL}&=\sum_{p_F<k<\Lambda} \varepsilon_k
(c^{\dag}_{\veck,\tau} c_{\veck,\tau}+d^{\dag}_{\veck,\tau}
d_{k,\tau})
\nonumber
\\&&
-2G_s\sum_{p_F<|\veck|,|\veck'|<\Lambda}\left[
{|\veck|\over\varepsilon_k}
(c^{\dag}_{\veck,\tau}
d^{\dag}_{-\veck,-\tau} -c_{\veck,\tau}d_{-\veck,-\tau})
{|\veck'|\over\varepsilon_k'}
(c^{\dag}_{\veck',\tau'}d^{\dag}_{-\veck',-\tau'}
-c_{\veck',\tau'}d_{-\veck',-\tau'})\right.
\nonumber\\
&&
+{}\left(c^{\dag}_{\veck,\tau}
d^{\dag}_{-\veck',-\tau'}
{T^{+}_{\tau,\tau'}}
-c_{\veck,\tau} d_{-\veck,-\tau'}
{T^{-}_{\tau,\tau'}}\right)
\left(c^{\dag}_{\veck',\tau''}
d^{\dag}_{-\veck',-\tau'''}
{T^{-}_{\tau'',\tau'''}}
-c_{\veck',\tau''} d_{-\veck',-\tau'''}
{T^{+}_{\tau'',\tau'''}}\right)
\nonumber\\
&&
+{}\left(c^{\dag}_{\veck,\tau}
d^{\dag}_{-\veck,-\tau'}
{T^{-}_{\tau,\tau'}}
-c_{\veck,\tau} d_{-\veck,-\tau'}
{T^{+}_{\tau,\tau'}}\right)\left(c^{\dag}_{\veck',\tau''}
d^{\dag}_{-\veck',-\tau'''}
{T^{+}_{\tau'',\tau'''}}
-c_{\veck',\tau''} d_{-\veck',-\tau'''}
{T^{-}_{\tau'',\tau'''}}\right)
\nonumber\\
&&+\left.{2}\left(c^{\dag}_{\veck,\tau}
d^{\dag}_{-\veck,-\tau'}
{T^{0}_{\tau,\tau'}}
-c_{\veck,\tau} d_{-\veck,-\tau'}
{T^{0}_{\tau,\tau'}}\right)\left(c^{\dag}_{\veck',\tau''}
d^{\dag}_{-\veck',-\tau'''}
{T^{0}_{\tau'',\tau'''}}
-c_{\veck',\tau''} d_{-\veck',-\tau'''}
{T^{0}_{\tau'',\tau'''}}\right)\right]\,.
\nonumber
\end{eqnarray}
The following commutators are easily obtained
\begin{eqnarray}
&[{\cal H}^{(2)}_{NJL},c^{\dag}_{\veck,\tau}d^{\dag}_{-\veck,-\tau'}T^{+}_{\tau,\tau'}]
&\approx 2\varepsilon_k
c^{\dag}_{\veck,\tau}d^{\dag}_{-\veck,-\tau'}
T^{+}_{\tau,\tau'}
\nonumber\\&&-{16G_s}
\sum_{p_F<|\veck'|<\Lambda}
\left(c^{\dag}_{\veck',\tau''}
d^{\dag}_{-\veck',-\tau'''}
{T^{+}_{\tau'',\tau'''}}
-c_{\veck',\tau''} d_{-\veck',-\tau'''}
{T^{-}_{\tau'',\tau'''}}\right)\,,
\nonumber
\\
&[{\cal H}^{(2)}_{NJL},c_{\veck,\tau}d_{-\veck,-\tau'}T^{-}_{\tau,\tau'}]
&\approx -2\varepsilon_k
c_{\veck,\tau}d_{-\veck,-\tau'}T^{-}_{\tau,\tau'}\nonumber\\&&-{16G_s}
\sum_{p_F<|\veck'|<\Lambda} \left(c^{\dag}_{\veck',\tau''}
d^{\dag}_{-\veck',-\tau'''}
{T^{+}_{\tau'',\tau'''}}
-c_{\veck',\tau''} d_{-\veck''',-\tau'}
{T^{-}_{\tau'',\tau''}}\right)\,.
\nonumber
\end{eqnarray}
Thus
\begin{eqnarray}
&&[{\cal H}^{(2)}_{NJL},\sum_{p_F<|\veck|<\Lambda}{1\over2\varepsilon_k}
(c^{\dag}_{\veck,\tau}d^{\dag}_{-\veck,-\tau'}T^{+}_{\tau,\tau'}
+c_{\veck,\tau}d_{-\veck,-\tau'}T^{-}_{\tau,\tau'})]
\nonumber\\&&\approx 
\left(1-\sum_{p_F<|\veck'|<\Lambda}{16G_s\over\varepsilon_{k'}}\right)
\sum_{p_F<|\veck|<\Lambda}
\left(c^{\dag}_{\veck,\tau}
d^{\dag}_{-\veck',-\tau'}
{T^{+}_{\tau,\tau'}}
-c_{\veck,\tau} d_{-\veck,-\tau'}
{T^{-}_{\tau,\tau'}}\right)=0.
\nonumber
\end{eqnarray}
This is the pion mode.
Notice that the NJL gap equation reads
$$1-\sum_{p_F<|\veck|<\Lambda}{16G_s\over\varepsilon_{k}}
=0$$
and $G_s>0.$

We present
next the computation of scalar modes.
The following commutators are easily obtained
\begin{eqnarray}
&[{\cal
H}^{(2)}_{NJL},c^{\dag}_{\veck,\tau}d^{\dag}_{-\veck,-\tau}]
&\approx 2\varepsilon_k
c^{\dag}_{\veck,\tau}d^{\dag}_{-\veck,-\tau}
-16G_s\,{|\veck|\over\varepsilon_k}\sum_{p_F<|\vec
k'|<\Lambda}{|\veck'|\over\varepsilon_{k'}}
(c^{\dag}_{\veck',\tau'}d^{\dag}_{-\veck',-\tau'}
-c_{\veck',\tau'} d_{-\veck',-\tau'})\,, \nonumber
\\
&[{\cal H}^{(2)}_{NJL},c_{\veck,\tau}d_{-\veck,-\tau}]
&\approx -2\varepsilon_k
c_{\veck,\tau}d_{-\veck,-\tau}
-16G_s\,{|\veck|\over\varepsilon_k}\sum_{p_F<|\veck'|<\Lambda}
{|\veck'|\over\varepsilon_{k'}}
(c^{\dag}_{\vec k',\tau'}d^{\dag}_{-\vec k',-\tau'}
-c_{\veck',\tau'} d_{-\veck',-\tau'})\,.
\nonumber\end{eqnarray}
It follows that
\begin{eqnarray}&&\left[{\cal H}^{(2)}_{NJL},\sum_{p_F<|\veck|<\Lambda}
\left(X_{\veck}c^{\dag}_{\veck,\tau}d^{\dag}_{-\veck,-\tau}
+Y_{\veck} c_{\veck,\tau}d_{-\veck,-\tau}\right)
\right]\nonumber
\approx\sum_{p_F<|\veck|<\Lambda}2\varepsilon_k
(X_{\veck}c^{\dag}_{\veck,\tau}d^{\dag}_{-\veck,-\tau}
-Y_{\veck}c_{\veck,\tau}d_{-\veck,-\tau}
)\nonumber\\
&&-16G_s\sum_{p_F<|\veck|<\Lambda}{(X_{\veck}+Y_{\veck})}
{|\veck|\over\varepsilon_k}
\sum_{p_F<|\veck'|<\Lambda}
{|\veck'|\over\varepsilon_{k'}}
(c^{\dag}_{\veck',\tau'}d^{\dag}_{-\veck',-\tau'}
-c_{\veck',\tau'}d_{-\veck',-\tau'}
)\nonumber
\nonumber\\
&&=\Omega_s
\sum_{p_F<|\veck|<\Lambda}
\left(X_{\veck}c^{\dag}_{\veck,\tau}d^{\dag}_{-\veck,-\tau}
+Y_{\veck} c_{\veck,\tau}d_{-\veck,-\tau}\right)\,.
\nonumber
\end{eqnarray}
The RPA equations read
\begin{eqnarray}&&2\varepsilon_k
X_{\vec k}-16G_s
{|\veck|\over\varepsilon_k}\sum_{p_F<|\veck'|<\Lambda}{(X_{\veck'}+Y_{\veck'})
} {|\veck'|\over\varepsilon_{k'}}=\Omega_s
X_{\veck}\,,\nonumber
\\
&&2\varepsilon_k
Y_{\veck}-16G_s
{|\veck|\over\varepsilon_k}\sum_{p_F<|\veck'|<\Lambda}{(X_{\veck'}+Y_{\veck'})
} {|\veck'|\over\varepsilon_{k'}}=-\Omega_sY_{\veck}\,,\nonumber
\end{eqnarray}
where $X_{\veck},Y_{\veck}$ are the RPA amplitudes and $\Omega_s$
the renormalized sigma mass.
The following dispersion relation is
obtained
$$1=
-16G_s\sum_{p_F<|\veck|<\Lambda}
{k^2\over\varepsilon_k^2}{4\varepsilon_k\over\Omega^2-4\varepsilon_k^2}
$$
showing that $\Omega_s=2M$. Indeed, replacing $\Omega^2$ by $4M^2$
in the l.h.s. of the dispersion equation
we obtain an identity, in view of the NJL gap equation,
\begin{eqnarray}&1&=
-16G_s\sum_{p_F<|\veck|<\Lambda}
{k^2\over\varepsilon_k^2}{4\varepsilon_k\over 4m^2-4\varepsilon_k^2}\nonumber\\
&&=
-16G_s\sum_{p_F<|\veck|<\Lambda}
{k^2\over\varepsilon_k^2}{4\varepsilon_k\over -4k^2}=1\,.\nonumber
\end{eqnarray}
In conclusion, we have shown that both models generate
pseudoscalar modes with zero mass. However, the sigma model generates a
scalar mode with mass close to $m_\sigma$ while the NJL model for nucleons generates a
scalar mode with an unacceptable mass, equal to $2M.$


\begin{thebibliography}{999}

\bibitem{2)} A. Chodos, R.L. Jaffe, K. Johnson, C.B. Thorn
and V.F. Weisskopf, {\it Phys. Rev.} {\bf D 9}  (1974) 3471.
\bibitem{hooft} G. 't Hooft, {\it Phys. Rev.} {\bf  D 14} (1976) 3432.
\bibitem{Klevansky92} S. P. Klevansky, {\it Rev. Mod. Phys.} {\bf64} (1992) 649.
\bibitem{PRS} J. da Provid\^encia, M.C. Ruivo and C.A. de Sousa {\it Phys. Rev.} {\bf D36} (1987) 1882.
\bibitem{4)} J. Carlson, K. Kogut and V.R. Pandharipande, {\it Phys. Rev.} {\bf D 27} (1983) 233;
 R. Sommer and J. Wosiek, {\it  Nucl. Phys.} {\bf B 267} (1986) 531 .
\bibitem{0)} P.A.M. Guichon, Phys. Lett B 200 (1988) 235.
\bibitem{01)} S. Fleck, W. Benz, K. Shimizu and K. Yazaki, Nucl.
Phys. {\bf A 510} (1990) 731-739.
\bibitem{Uehara} M. Uehara and H. Kondo, {\it Prog. Theor. Phys.} {\bf 71} (1984) 1303.
\bibitem{walecka} B.D. Serot and J.D. Walecka, {\it Adv. Nucl. Phys.} {\bf 16}
(1986) 1; {Int. J. Mod. Phys. E} {\bf 16} (1997) 515.
\bibitem{Luso} H. Bohr,
Proceedings of the International Workshop "Fission Dynamics of Atomic Clusters and Nuclei",
Eds. J. da Provid\^encia, D. Brink, F.F. Karpeshin and F. B. Malyk,
World Scientific, Singapore, (2001) 65.
\bibitem{1)} H.R. Petry, H. Hofest\"adt,
S. Merk, K. Bleuler, H. Bohr and K.S. Narain,  {\it Phys. Lett.} {\bf 159 B} (1985) 363 .
\bibitem{Moszk} S. A. Moszkowski, private communication with J.P.
\bibitem{buervenich} T.J. B\"uvernich and D.G. Madland, {\it Nuclear Physics} {\bf A 729} (2003) 769.
\bibitem{Koch87} V. Koch, T.S. Biro, J. Kunz, and U. Mosel,  Phys. Lett. {\bf B185},\,1\, (1987).
\bibitem{ring} G.A. Lalazissis, J. K\"onig, P. Ring, Phys. Rev. {\bf C55},\,  540\, (1997).
\bibitem{chanfray} G. Chanfray, M. Ericson and P.A.M. Guichon, {\it Phys. Rev.} {\bf C 63} (2001) 055202;
 G. Chanfray, M. Ericson, arXiv:nucl-th/0402018.
\bibitem{Glend} Normann K. Glendenning, "Compact Stars",  Springer, (1996)
\end{thebibliography}
\end{document}